\documentclass[conference]{IEEEtran}
\IEEEoverridecommandlockouts

\usepackage{cite}
\usepackage{amsmath,amssymb,amsfonts}
\usepackage{algorithmic}
\usepackage{graphicx}
\usepackage{textcomp}
\usepackage[utf8]{inputenc}
\DeclareUnicodeCharacter{202F}{\,}
\usepackage{xcolor}
\def\BibTeX{{\rm B\kern-.05em{\sc i\kern-.025em b}\kern-.08em
    T\kern-.1667em\lower.7ex\hbox{E}\kern-.125emX}}
\begin{document}

\title{Long-Range Rendezvous and Docking Maneuver Control of Satellite using Cross-Feedback Sliding Mode Controller\\
\thanks{\\
† Joint first authors\\\hspace{10cm}
* Corresponding author: rakesh.sahoo266@gmail.com}
}

\author{
\IEEEauthorblockN{Vedant Vivek Kini\textsuperscript{†}} 
\IEEEauthorblockA{\textit{Department of Aerospace Engineering} \\
\textit{Indian Institute of Technology Kharagpur}\\
West Bengal, 721302, India\\
vedantkini21@kgpian.iitkgp.ac.in}
\\
\hspace{1.5cm}
\IEEEauthorblockN{Rakesh Kumar Sahoo\textsuperscript{*}}  
\IEEEauthorblockA{\textit{Department of Aerospace Engineering} \\
\textit{Indian Institute of Technology Kharagpur}\\
West Bengal, 721302, India\\
rakesh.sahoo266@gmail.com}
\and  
\IEEEauthorblockN{Dantu Phani Surya\textsuperscript{†}}  
\IEEEauthorblockA{\textit{Department of Aerospace Engineering} \\
\textit{Indian Institute of Technology Kharagpur}\\
West Bengal, 721302, India\\
suryadantu18@kgpian.iitkgp.ac.in}
\\
\hspace{1.5cm}
\IEEEauthorblockN{Manoranjan Sinha}  
\IEEEauthorblockA{\textit{Department of Aerospace Engineering} \\
\textit{Indian Institute of Technology Kharagpur}\\
West Bengal, 721302, India\\
masinha@aero.iitkgp.ac.in}
}

\maketitle

\begin{abstract}
Satellite rendezvous and docking (RvD) maneuvers are essential for satellite servicing and in-orbit assembly. Traditional approaches often treat translational and rotational motions independently, simplifying control design but potentially leading to inefficiencies in maneuver time and fuel consumption. To address these challenges, a novel cross-feedback sliding mode controller has been proposed, developing an interdependent regulation system for translational and rotational motion. This method decouples the relative translational and rotational motion of chaser satellite with respect to target satellite while incorporating cross-feedback mechanisms to account for their inherent coupling. By incorporating rotational state information into translational control laws and vice versa, the approach ensures coordinated adjustments, enhancing maneuver efficiency.  The chaser satellite manages both translational and rotational adjustments to rendezvous and dock with the target satellite. The stability of the cross-feedback sliding mode controller is established within the Lyapunov framework, and simulation results substantiate the effectiveness of this strategy.
\end{abstract}


\section{Introduction}
Satellite rendezvous and docking (RvD) maneuvers are among the most intricate and mission-critical operations, requiring precise coordination between two satellite to achieve successful alignment and docking \cite{b1}. The rendezvous maneuver involves bringing the chaser satellite closer to the target \cite{b2}, \cite{b3}, \cite{b4}. This is followed by the docking phase, where precise alignment and a secure physical connection between their docking ports are achieved by rotating the chaser satellite or both satellites, depending on feasibility \cite{b5}, \cite{b6}. These maneuvers are fundamental to a wide range of space applications, including crewed spaceflight, orbital resupply missions, satellite servicing, and the in-orbit assembly of large space structures \cite{b7}. The significance of RvD is exemplified by missions such as ISRO’s SpaDeX, which successfully demonstrated India’s ability to autonomously dock two satellites in orbit. Such advancements underscore the critical role of precise guidance, navigation, and control strategies in ensuring mission success. 

Traditionally, satellite RvD maneuvers have been designed by treating translational and rotational motion as independent processes \cite{b8}. Translational motion governs the satellite’s trajectory in three-dimensional space, actuated using the thruster to regulate velocity and position. Conversely, rotational motion, or attitude dynamics, determines the satellite’s orientation, actuated using reaction wheels and control moment gyroscope to generate the required torques. This decoupled approach simplifies the controller design and also reduces the computational requirement \cite{b9}. For instance, during the Apollo missions, the Lunar and Command Modules executed RvD maneuvers by sequentially adjusting position and orientation, reducing control complexity \cite{b10}. However, this stepwise approach resulted in increased maneuver duration and fuel consumption, as translational and rotational adjustments were performed separately. Given the stringent fuel constraints in space missions, such inefficiencies could impose operational limitations or even jeopardize mission success. Despite its conceptual simplicity, the independent treatment of translation and rotation overlooks the inherent coupling between these dynamics. In reality, satellite attitude changes influence trajectory by altering thrust direction and external force interactions. Neglecting these interdependencies can degrade maneuver accuracy, requiring additional corrections, increasing fuel consumption, and prolonging mission duration. 

To address the limitations of the traditional decoupled approach and ensure robust and efficient RvD maneuvers, numerous studies in the literature have formulated translational and rotational dynamics within a unified control framework \cite{b11}, \cite{b12} in close proximity of target satellite. Coupled motion control facilitates simultaneous adjustments in position and orientation, leading to improved maneuver efficiency by minimizing both time and fuel consumption. This integrated approach is particularly advantageous for close-proximity RvD operations \cite{b13}, \cite{b14}, \cite{b15}. However, the incorporation of coupled dynamics significantly increases the complexity of the mathematical model, making real-time implementation computationally intensive. Moreover, improper tuning of the control parameters may induce instability or unintended system behavior, as errors in one motion component can propagate and amplify, potentially compromising the overall maneuver performance. 

In this paper, we have proposed a novel method for efficient long-range rendezvous and docking maneuvers to address the limitations of both decoupled and coupled approaches described above. This approach decouples the translational and rotational motion of each satellite, as in traditional decoupled methods, but introduces a cross-feedback sliding mode controller \cite{b16} to account for the inherent coupling between these dynamics. Unlike conventional decoupled control, cross-feedback control establishes an interdependent regulation system by incorporating rotational state information into the translational control law and vice versa. This ensures that corrections in one domain (translational or rotational) are compensated by adjustments in the other, enabling a more coordinated and efficient maneuver. In this hybrid approach, the chaser satellite performs both translational and rotational motion, while the target satellite executes only rotational motion, optimizing the overall maneuver process.

The remainder of the paper is structured as follows. Section 2 outlines the rigid body dynamics of both the chaser and target satellites, followed by the relative kinematics and dynamics of the chaser with respect to the docking port of the target satellite. Section 3 introduces the proposed cross-feedback sliding mode control scheme. Finally, Section 4 concludes with the simulation results, demonstrating the effectiveness of the proposed approach. 

\section{Rigid Body Dynamics of Satellites}
This section presents the rigid body dynamic model of the target and the chaser satellite orbiting around the Earth. To model the dynamics three reference frames are defined as International Celestial Reference Frame (ICRF) \cite{b17} denoted by  $E_i$., the chaser body-fixed frame denoted by \( E_c \) and the target body-fixed frame denoted by \(E_t\). 
\subsection{Rigid Body Dynamics and Kinematics of Target Satellite}\label{AA}
Let the position vector of  the target satellite with respect to $E_t$ be denoted by $\tilde{p}_t$, whereas the translational and angular velocity of target satellite w.r.t $E_i$ frame be denoted by $\tilde{v}_t$ and $\tilde{\omega}_t$, respectively. The translational kinematics and dynamics of the target satellite can be described as \cite{b18}
\begin{equation}
M_t\dot{\tilde{v}}_t +M_t \mathbf{\Omega}(\tilde{\omega}_t) \tilde{v}_t\,=\,\tilde{F}_{dt}+\tilde{F}_{j_2,t}\label{target_trans_dyn}
\end{equation}
\begin{equation}
    \dot{\tilde{p}}_t +\mathbf{\Omega}(\tilde{\omega}_t) \tilde{p}_t\,=\, \tilde{v}_t 
\end{equation}
where \(M_t\) is the mass of the target satellite, taken as \(1000\ kg\), \(\tilde{F}_{d}\) is the external disturbance force and \(\tilde{F}_{j_2,t}\)  is gravity and \(J_2\) perturbation force acting on the target satellite \cite{b18}, respectively. \(\mathbf{\Omega}(\tilde{\omega}_t)\) denotes the skew-symmetric matrix representation of angular velocity vector $\tilde{\omega}_t)$ as defined below.
\begin{equation*}
    \mathbf{\Omega({\tilde{\omega}_t})} = 
    \begin{bmatrix} 
        0 & -\omega_{t_z} & \omega_{t_y} \\ 
        \omega_{t_z} & 0 & -\omega_{t_x} \\ 
        -\omega_{t_y} & \omega_{t_x} & 0 
    \end{bmatrix}
\end{equation*}
The rotational kinematics and dynamics of the target satellite is described as given below.
\begin{equation}
    \mathbf{J_t} \dot{\tilde{\omega}}_t = -\mathbf{\Omega}({\tilde{\omega}}_t) \mathbf{J_t} \tilde{\omega}_t + \tilde{T}_{dt}+\tilde{T}_{gt}
    \label{target_rot_dyn}
\end{equation}
where  \(\tilde{T}_{dt}\) represent the external disturbance torque and \(\tilde{T}_{gt}\) denote the gravity gradient torque acting on the target satellite\cite{b1}; \( \textbf{J}_t \in \mathbb{R}^{3 \times 3} \) denotes the inertia matrix of the target satellite. 
The attitude kinematics of the target satellite is given below:
\begin{equation}
\dot{\boldsymbol{B}}_{t}^{i} = \boldsymbol{B}_{t}^{i}\, \mathbf{\Omega}({\tilde{\omega}}_t)
\label{target_rot_kinematics}
\end{equation}
where  $\boldsymbol{B}_{t}^{i}\in\,SO(3)$ is the rotation matrix which transforms from the target body-fixed frame to ICRF frame.

\subsection{Rigid Body Dynamics and Kinematics of Chaser Satellite}
Let the position vector of  the chaser satellite with respect to $E_c$ be denoted by $\tilde{p}_c$, whereas the translational and angular velocity of chaser satellite w.r.t $E_i$ frame be denoted by $\tilde{v}_c$ and $\tilde{\omega}_c$, respectively. The kinematics and dynamics of the translational motion of the chaser satellite can be described as 
\begin{equation}
M_c \dot{\tilde{v}}_c +M_c \mathbf{\Omega}(\tilde{\omega}_c) \tilde{v}_c  =\tilde{F}_{dc}+\tilde{F}_{j_2,c}+\tilde{F}_c\label{chaser_trans_dyn}
\end{equation}
\begin{equation}
    \dot{\tilde{p}}_c +\mathbf{\Omega}({\tilde{\omega}}_c) \tilde{p}_c\,=\, \tilde{v}_c 
\end{equation}
where \(M_c\) is the mass of the chaser satellite, \(\tilde{F}_{dc}\), \(\tilde{F}_{c}\) and \(\tilde{F}_{j_2,c}\) are the external disturbance force, control input force and \(J_2\) perturbation force acting on the chaser satellite. The attitude dynamics and kinematics of the chaser satellite are given by:
\begin{equation}
    \textbf{J}_c \dot{\tilde\omega}_c = -\mathbf{\Omega}({\tilde\omega}_c) \textbf{J}_c {\tilde\omega}_c + \tilde{T}_{dc}+T_{gc}+ \tilde{T}_{c}
    \label{chaser_rot_dyn}
\end{equation}
where \(\mathbf{J_c}\) represents the inertia matrix of the chaser satellite, \(\tilde{T}_{dc}\), \(\tilde{T}_{c}\) and \(\tilde{T}_{gc}\) are the external disturbance torque, control input torque and gravity-gradient torque acting on the chaser satellite \cite{b18}. The attitude kinematics of the chaser satellite is given below:
\begin{equation}
\dot{\boldsymbol{B}}_{c}^{i} = \boldsymbol{B}_{c}^{i}\, \mathbf{\Omega}({\tilde{\omega}}_c)
\label{chaser_rot_kinematics}
\end{equation}
where  $\boldsymbol{B}_{c}^{i}\in\,SO(3)$ is the rotation matrix which transforms from the chaser body-fixed frame to ICRF frame.
 \begin{figure}
    \centering
    \includegraphics[width=1\linewidth]{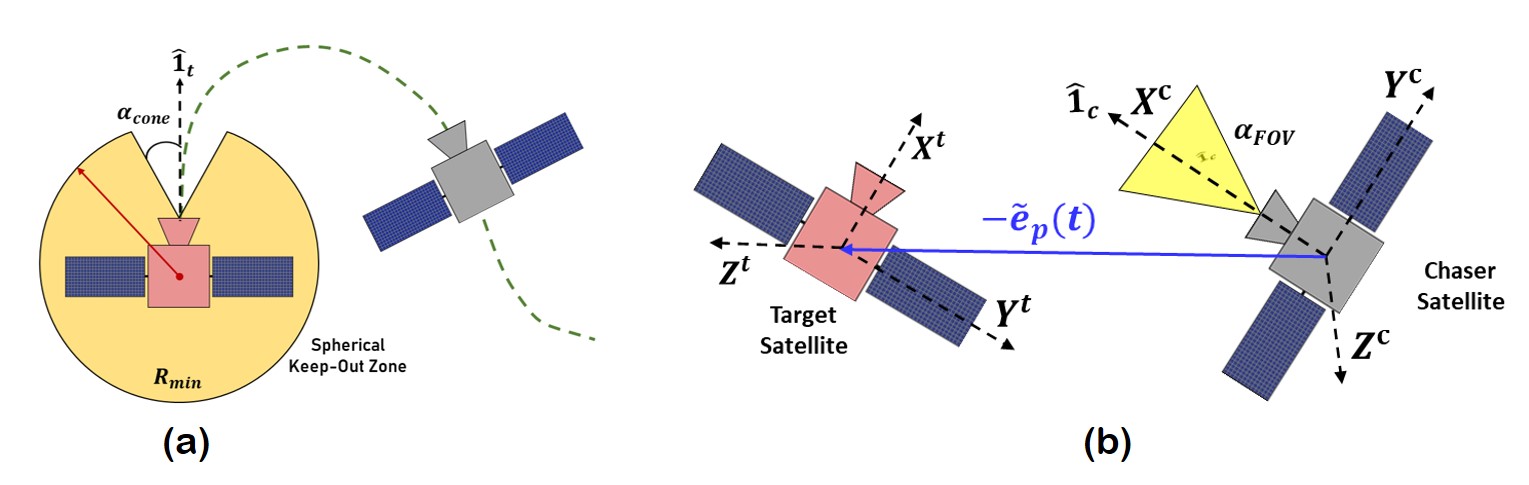}
    \caption{Illustration of trajectory constraints for precise rendezvous and docking maneuver}
    \label{constraints}
\end{figure}
\subsection{Relative Kinematics and Dynamics of Satellites}
Let the relative translational position and the relative translational velocity  of the chaser satellite with respect to the target satellite is given by $\tilde{e}_p = \tilde{p}_c - \mathbf{B_t^c} \tilde{p}_t$ and  $\tilde{e}_{v} = \tilde{v}_c - \mathbf{B_t^c} \tilde{v}_t$, respectively. The relative acceleration of chaser satellite with respect to the target satellite can described as shown below.
\begin{equation*}
\begin{aligned}\dot{\tilde{e}}_{v}\,=\, & \frac{\tilde{F}_{c}}{M_{c}}+\frac{\tilde{F}_{j_{2},c}}{M_{c}}+\frac{\tilde{F}_{dc}}{M_{c}}+\left(\mathbf{\Omega}\left(\tilde{e}_{\omega}\right)+\mathbf{\Omega}\left(\tilde{\omega}_{t}\right)\right)B_{t}^{c}\tilde{v}_{t}\\
 &\, -\mathbf{\Omega}\left(\tilde{\omega}_{c}\right)\tilde{v}_{c}-\textbf{B}_{t}^{c}\frac{\tilde{F}_{dt}}{M_{t}}
\end{aligned}
\end{equation*}
Relative translational dynamics is obtained by multiplying the above equation with the mass of the chaser satellite \(M_c\) yields
\begin{equation}
\begin{aligned}M_{c}\dot{\tilde{e}}_{v}+M_{c}\mathbf{\Omega}\left(\tilde{\omega}_{c}\right)\tilde{v}_{c}\,=\, & \tilde{F}_{c}+\tilde{F}_{j_{2},c}+\tilde{F}_{dc}-M_{c}B_{t}^{c}\frac{\tilde{F}_{dt}}{M_{t}}+\\
 &\,M_{c}\left(\mathbf{\Omega}\left(\tilde{e}_{\omega}\right)+\mathbf{\Omega}\left(\tilde{\omega}_{t}\right)\right)B_{t}^{c}\tilde{v}_{t} 
\end{aligned}
\label{trans_rel_dyn}
\end{equation}
The relative attitude tracking error of chaser satellite w.r.t target satellite is given by $\tilde{e}_r=log([\boldsymbol{B}_{t}^{i}]^T\,[\boldsymbol{B}_{c}^{i}])$, where $log(.)$ is the logarithmic mapping from $SO(3)$ to $\mathbb{R}^{3}$. The relative angular velocity vector of chaser satellite w.r.t target satellite is expressed as $\tilde{e}_\omega = \tilde\omega_c - \mathbf{B_t^c} \tilde\omega_t$, where $\mathbf{B_t^c} \in \mathbb{R}^{3 \times 3}$ denotes the rotation matrix from the $E_t$ to $E_c$ and is formulated as $\mathbf{B_t^c}=\left(\mathbf{B_c^i}\right)^T\mathbf{B_t^i}$. The relative angular acceleration is given by performing time-derivative as shown below.
\begin{equation*}
\dot{\tilde{e}}_\omega \,=\, \dot{\tilde\omega}_c + \tilde{\omega} \times \mathbf{B_d^c} \omega_d - \mathbf{B_d^c} \dot{\tilde{\omega}}_d
\end{equation*}
where subscript $(.)_d$  represents the desired position or orientation to be achieved by the chaser satellite at the end of docking phase. Relative rotational dynamics is obtained by multiplying the above equation by \(\textbf{J}_c\) as shown below.
\begin{equation}
\mathbf{J_c}\dot{\tilde{e}}_\omega \,=\, \mathbf{J_c}\dot{\tilde\omega}_c + \mathbf{J_c}\,\tilde{\omega} \times \mathbf{B_d^c} \omega_d - \mathbf{J_c}\mathbf{B_d^c} \dot{\tilde{\omega}}_d
\label{rot_rel_dyn}
\end{equation}
  \begin{figure}
    \centering
    \includegraphics[width=1\linewidth]{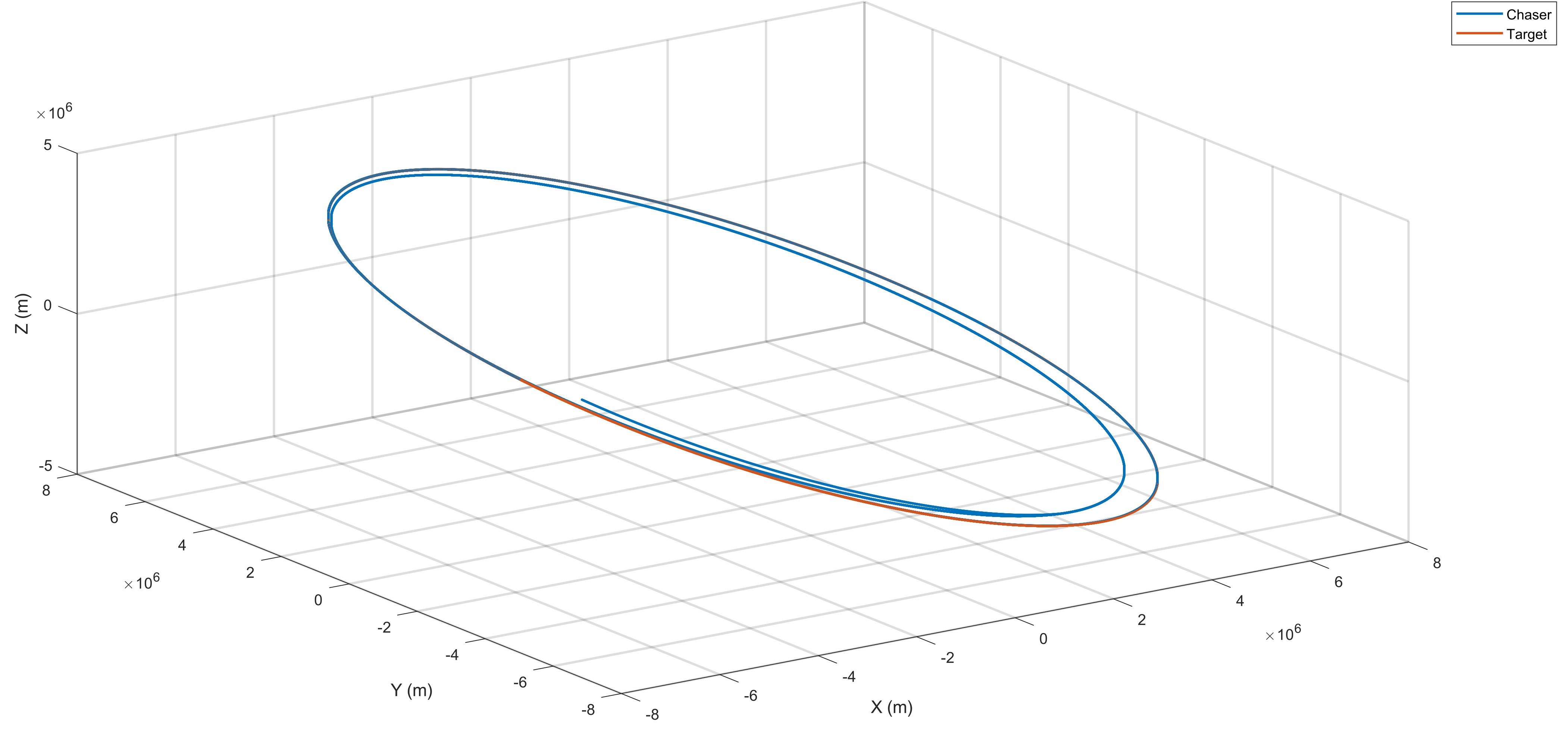}
    \caption{Plot of 3D trajectory of chaser satellite and target satellite in ICRF Frame during rendezvous maneuver}
    \label{3d_trajectory}
\end{figure}
\section{Trajectory Optimization}
The trajectory optimization framework is designed to first obtain an optimal state vector for translational motion while satisfying mission-critical constraints. The propellant consumption is predominantly governed by thruster firings, which directly translate into changes in linear velocity, $\triangle v$. This makes $\triangle v$ a practical and mission-relevant metric for estimating fuel expenditure. The rotational control effort is not neglected in the overall design. It is fully integrated into the control strategy via the cross-feedback sliding mode controller (CxSMC), where both translational and rotational states are simultaneously regulated. Moreover, the attitude maneuver of the chaser satellite is constrained by the field-of-view constraint as explained below.
The optimization problem aims to minimize the chaser satellite's impulse via the objective function: 
\begin{equation*}
    J \,=\,minimize \left(\sum^n_{i=1} \triangle v_i\right)^2
\end{equation*}
subject to nonlinear inequality constraints: \\
\\
(a) \textbf{Spherical keep-out zone}$\,:\,\left\Vert \tilde{e}_{p}\right\Vert \,\geq\,R_{min}$\\
Maintains a minimum distance $R_{min}$ from the target to prevent collisions as shown in Fig. \ref{constraints}(a). \\
\\
(b) \textbf{Approach conic corridor}$\,:\,\left\Vert \tilde{e}_{p}\right\Vert cos(\alpha_{cone})\leq\hat{1}_{t}\tilde{e}_{p}$\\
Confines the chaser within a cone (half-angle $\alpha_{cone}$ ) aligned with the target’s docking axis $\hat{1}_{t}$ during terminal approach as illustrated in Fig. \ref{constraints}(a). \\
\\
(c) \textbf{Field of view constraint}$:\,\left\Vert \tilde{e}_{FOV}\right\Vert cos(\alpha_{FOV})\leq\hat{1}_{c}\tilde{e}_{FOV}$
Ensures continuous target visibility within the chaser’s sensor cone half-angle $\alpha_{FOV}$ along its docking axis $\hat{1}_{c}$ as depicted in Fig. \ref{constraints}(b), where  $\tilde{e}_{FOV}$ is the relative position vector of target satellite w.r.t chaser satellite.\\
\\
(e) \textbf{Velocity Constraint}: This constraint enforce a constant velocity of $0.3\,m/s$ during mid-range proximity maneuver from $1\,km$ to $10\,m$ and reduces it to  $0.03\,m/s$ for final soft docking.
\begin{equation*}
    \begin{aligned}\tilde{e}_{v} & =\left\{ \begin{array}{cc}
0.3m/s & ,if\,\,10m<\left\Vert \tilde{e}_{p}\right\Vert \leq1000m\\
0.03m/s & ,if\,\,\left\Vert \tilde{e}_{p}\right\Vert \leq10m\qquad\qquad
\end{array}\right.\end{aligned}
\end{equation*}
\\
(e) \textbf{Impulse Constraint}: Since the maneuver is impulsive in nature, the focus is on minimizing the change in velocity ($\triangle v$) rather than the applied force, as the force is not explicitly part of the state vector. The control inputs are modeled as instantaneous impulses, making $\triangle v$ the primary metric of interest for optimization.

\begin{equation*}
-300\sqrt{3} < \left\Vert \triangle v \right\Vert < 300\sqrt{3}
\end{equation*}
The trajectory optimization framework is designed to act as higher-level guidance layer generating feasible trajectory of chaser satellite. As $\triangle v$ represents change in velocity at discrete time step due to impulsive thrust. These impulses are optimized to minimize the total velocity change from initial state to the desired docking state. The optimized impulse $\triangle v$ is used to compute discrete-time state propagation of chaser satellite, $\tilde{p}_c(k+1)=\tilde{p}_c(k)+h[\boldsymbol{B}_{c}^{i}]^T\Delta v$ where $h=t_{k+1}-t_{k}$ is fixed-time step size.

\section{Design of Cross-Feedback Sliding Mode Controller}

In this section, we have implemented a cross-feedback sliding mode controller (CxSMC) to track the trajectory for the chaser satellite in the three-dimensional configuration space SE(3). Stability of the controller is established using a Lyapunov framework. 
\\
\textbf{Lemma 1}: For a nonlinear system expressed as $\dot{\tilde{y}}=f\left(\tilde{y}\right)$, $\tilde{y}\left(0\right)=\tilde{y}_{0}$, $\tilde{y}\ \epsilon\ \mathbb{R}^{n}$ , if there exist a continuously differentiable Lyapunov function V(x) in the neighborhood $N\subset\mathbb{R}^{n}$ of origin, which satisfies $\dot{V}\left(\tilde{y}\right)\le\ -kV\left(\tilde{y}\right)$ where $k>0$. Then the origin of the system is asymptotically stable and the settling time T(y) is dependent on initial states as shown below.
\begin{equation*}
    T\left(y\right)\le T_{max}=\frac{\ln(V)}{k}
\end{equation*}

\subsection{{Cross-Feedback Linear Sliding Surface}}
Instead of designing independent sliding surfaces for translational and rotational motion, the proposed cross-feedback sliding surface of translational motion consists of translational error and cross-feedback terms of rotational error and vice versa. The translational sliding surface denoted by $\tilde{s}_{p}$  \eqref{trans sliding surface} and the rotational sliding surface denoted by $\tilde{s}_{r}$ \eqref{rot sliding surface} is defined based on relative position $\tilde{e}_{p}$ and orientation error $\tilde{e}_{r}$   as shown below.
\begin{equation}
\tilde{s}_{p}=\dot{\tilde{e}}_{p}+\lambda_{p}\tilde{e}_{p}+\mu_{p}\tilde{e}_{r}
\label{trans sliding surface}
\end{equation}
\begin{equation}
    \tilde{s}_{r}=\dot{\tilde{e}}_{r}+\lambda_{r}\tilde{e}_{r}+\mu_{r}\tilde{e}_{p}
    \label{rot sliding surface}
\end{equation}
where $\lambda_{p},\lambda_{r}>0$ and $\mu_{p},\mu_{r}>0$ are cross-feedback gains. An exponential reaching law for both translational and rotational motion is defined to ensure that the system error states reaches the sliding surface as, $\dot{\tilde{s}}_{p}=-k_{p}\tilde{s}_{p}$  and $\dot{\tilde{s}}_{r}=-k_{r}\tilde{s}_{r}$
 \begin{figure}
    \centering
    \includegraphics[width=1\linewidth]{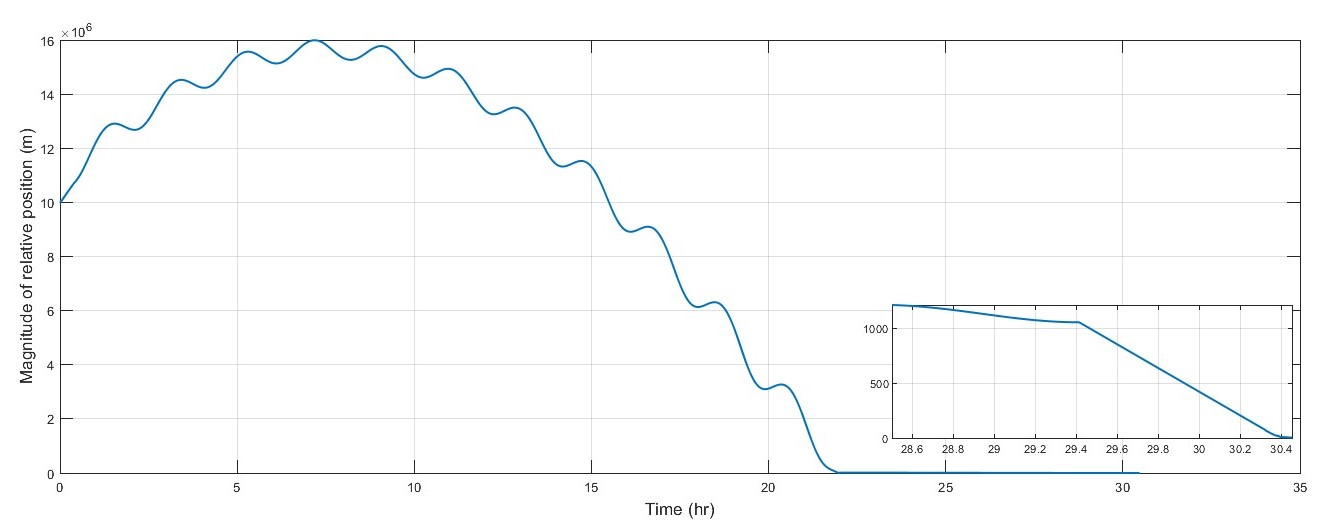}
    \caption{Relative position of chaser satellite with respect to target satellite}
    \label{rel_pos}
\end{figure}
\\
\textbf{Theorem 1}  (Stability of Reaching Phase) Considering the cross-feedback sliding surfaces as defined in \eqref{trans sliding surface} and \eqref{rot sliding surface} for the relative translational and rotational dynamics described in \eqref{trans_rel_dyn} and \eqref{rot_rel_dyn}, the relative position and velocity of the chaser satellite with respect to the docking port of the target satellite will converge to the sliding surface. The corresponding external control input in the form of force and torque applied to the chaser satellite is formulated as follows: 
\begin{equation}
 \begin{alignedat}{1}\tilde{F}_{c}\,=\, & -\tilde{F}_{j_{2},c}-\tilde{F}_{dc}-M_{c}\left(\boldsymbol{\Omega}\left(\tilde{e}_{\omega}\right)+\boldsymbol{\Omega}\left(\tilde{\omega}_{t}\right)\right)B_{t}^{c}\tilde{v}_{t}\\
 & \,+M_{c}\boldsymbol{\Omega}\left(\tilde{\omega}_{c}\right)\tilde{v}_{c}+B_{t}^{c}\left(\frac{M_{c}}{M_{t}}\right)\tilde{F}_{dt}-\lambda_{p}M_{c}\dot{\tilde{e}}_{p}\\
 & \,-\mu_{p}M_{c}\dot{\tilde{e}}_{r}-k_{p}\tilde{s}_{p}
\end{alignedat}
\label{control force}
\end{equation}
\begin{equation}
    \begin{alignedat}{1}\tilde{T}_{c}= & -\tilde{T}_{dc}-\tilde{T}_{gc}+\boldsymbol{\Omega}\left(\tilde{\omega}_{c}\right)\boldsymbol{J}_{c}\tilde{\omega}_{c}-\boldsymbol{J}_{c}\tilde{e}_{\omega}\times\boldsymbol{B}_{d}^{c}\dot{\tilde{\omega}}_{d}\\
 & +\boldsymbol{J}_{c}\boldsymbol{B}_{d}^{c}\dot{\tilde{\omega}}_{d}-\lambda_{r}\boldsymbol{J}_{c}\dot{\tilde{e}}_{r}-\mu_{r}\boldsymbol{J}_{c}\dot{\tilde{e}}_{p}-k_{r}\tilde{s}_{r}
\end{alignedat}
\label{control torque}
\end{equation}
\\
\textbf{Proof:}  Consider a continuous differentiable Lyapunov function as
 \begin{equation*}
     V_1= \frac{1}{2}\ \tilde{s}^{T}_p\mathbf{M}_{c}\tilde{s}_p \,\, +\,\, \frac{1}{2}\ \tilde{s}^{T}_r\mathbf{J}_{c}\tilde{s}_r
 \end{equation*} 
 The time derivative of Lyapunov function can be written using \eqref{control force} and \eqref{control torque}
  \begin{equation*}
\begin{aligned}\dot{V}_1 & =\tilde{s}_{p}^{T}M_c\dot{\tilde{s}}_{p}+\tilde{s}_{r}^{T}\mathbf{J}_{c}\dot{\tilde{s}}_{r}\\
 & =-k_{p}\tilde{s}_{p}^{T}M_c\tilde{s}_{p}-k_{r}\tilde{s}_{r}^{T}\mathbf{J}_{c}\tilde{s}_{r}\\
 & \leq-\lambda_{1}\left(\tilde{s}_{p}^{T}M_c\tilde{s}_{p}\right)-\lambda_{1}\left(\tilde{s}_{r}^{T}\mathbf{J}_{c}\tilde{s}_{r}\right)
\end{aligned}
 \end{equation*}
\begin{figure}
    \centering
    \includegraphics[width=1\linewidth]{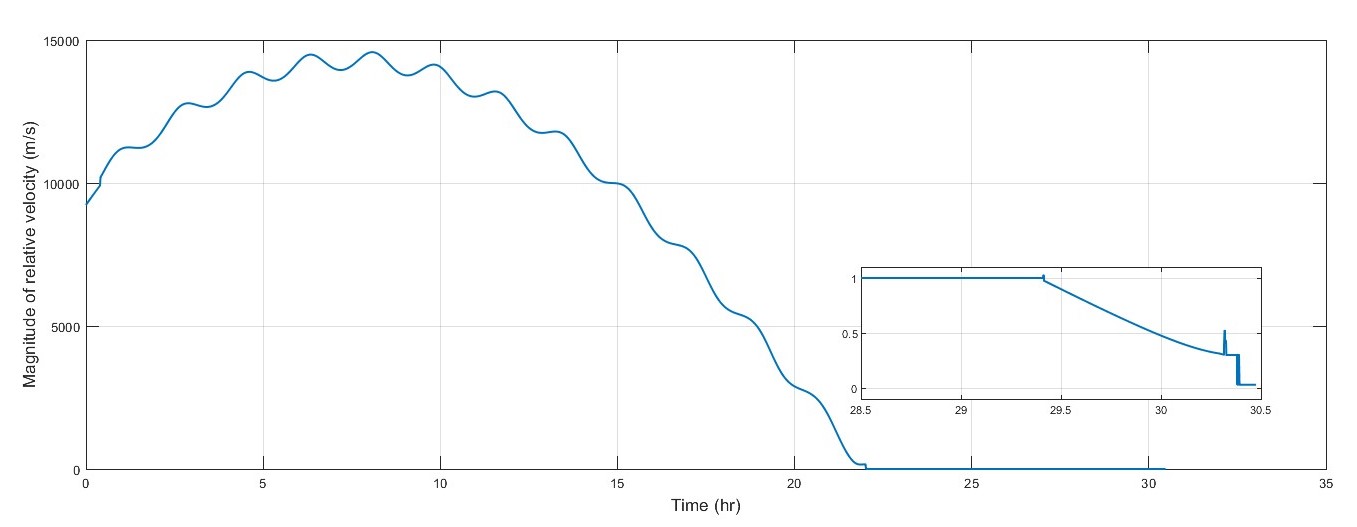}
    \caption{Relative velocity of chaser satellite with respect to the target satellite}
    \label{rel_vel}
\end{figure}
 where $\lambda_{1}=min\left(k_{p}\,\,,\,\,k_{r}\right)$ .  The Lyapunov derivative can be rewritten as
\begin{equation*}
    \begin{aligned}\dot{V}_{1} & \leq-\lambda_{1}\left(\tilde{s}_{p}^{T}\mathbf{M}_{c}\tilde{s}_{p}+\tilde{s}_{r}^{T}\mathbf{J}_{c}\tilde{s}_{r}\right)\\
 & \leq-\lambda_{1}\left(2V_{1}\right)\\
 & \leq-\lambda_{2}V_{1}
\end{aligned}
\end{equation*}
where  $\lambda_{2}=2\lambda_{1}$ and $\lambda_2>0$.  It can be inferred from the lyapunov derivative $\dot{V}_1\,\leq\,0$ that the system is asymptotically stable according to Lemma 1. Consequently, the error states converges to the desired sliding surface of translational and rotational motion $\tilde{s}_p=0$ and $\tilde{s}_r=0$  asymptotically. \\

\textbf{Theorem 2}  (Stability of Sliding Phase) Considering the cross-feedback sliding surface for translational and rotational motion as defined in \eqref{trans sliding surface} and \eqref{rot sliding surface} for the relative translational and rotational dynamics described in \eqref{trans_rel_dyn} and \eqref{rot_rel_dyn}, the relative position and velocity of the chaser satellite with respect to the target satellite will converge to equilibrium along the sliding surface asymptotically.
 \begin{equation*}
     T_s\left(x\right)\le \frac{\ln(V_2)}{\lambda_{4}}
 \end{equation*} \\
\textbf{Proof:}  Consider a continuous differentiable Lyapunov function as
  \begin{equation*}
     V_2= \frac{1}{2}\ \tilde{e}^{T}_p\tilde{e}_p \,\, +\,\, \frac{1}{2}\ \tilde{e}^{T}_r\tilde{e}_r
 \end{equation*} 
The time derivative of Lyapunov function is  
\begin{equation*}
 \begin{aligned}\dot{V}_2 & =\tilde{e}_{p}^{T}\dot{\tilde{e}}_{p}+\tilde{e}_{r}^{T}\dot{\tilde{e}}_{r}\end{aligned}
 \label{v_derivative_error}
\end{equation*}
when the error states evolve on the sliding surface i.e. $\tilde{s}_{p}=0$ and $\tilde{s}_{r}=0$, then $\dot{\tilde{e}}_{p}$ and $\dot{\tilde{e}}_{r}$ are given by $\dot{\tilde{e}}_{p}  =-\lambda_{p}\tilde{e}_{p}-\mu_{p}\tilde{e}_{r}$  and  $\dot{\tilde{e}}_{r} =-\lambda_{r}\tilde{e}_{r}-\mu_{r}\tilde{e}_{p}$ . The Lyapunov derivative can be rewritten  as
\begin{equation*}
\begin{aligned}\dot{V}_{2} & =-\tilde{e}_{p}^{T}\left(\lambda_{p}\tilde{e}_{p}+\mu_{p}\tilde{e}_{r}\right)-\tilde{e}_{r}^{T}\left(\lambda_{r}\tilde{e}_{r}+\mu_{r}\tilde{e}_{p}\right)\\
 &=-\lambda_{p}\tilde{e}_{p}^{T}\tilde{e}_{p}-\lambda_{r}\tilde{e}_{r}^{T}\tilde{e}_{r}-\left(\mu_{p}+\mu_{r}\right)\tilde{e}_{p}^{T}\tilde{e}_{r}\\
 & \leq-\lambda_{p}\tilde{e}_{p}^{T}\tilde{e}_{p}-\lambda_{r}\tilde{e}_{r}^{T}\tilde{e}_{r}\\
 & \leq-\lambda_{3}\left(\tilde{e}_{p}^{T}\tilde{e}_{p}+\tilde{e}_{r}^{T}\tilde{e}_{r}\right)\\
 & \leq-\lambda_{3}\left(2V_{2}\right)\\
\dot{V}_{2} & \leq-\lambda_{4}V_{2}
\end{aligned}
\end{equation*}
where $\lambda_{3}=min(\lambda_{p}\,,\,\lambda_{r})$,  $\lambda_{4}=2\lambda_{3}$ and $\lambda_4>0$.  It can be inferred from the lyapunov derivative $\dot{V}_2\,\leq\,0$ that the system is asymptotically stable by Lemma 1. Consequently, the error states converge to the equilibrium point along the desired sliding surface of translational and rotational motion $\tilde{s}_p=0$ and $\tilde{s}_r=0$ asymptotically.
 \begin{figure}
    \centering
    \includegraphics[width=1\linewidth]{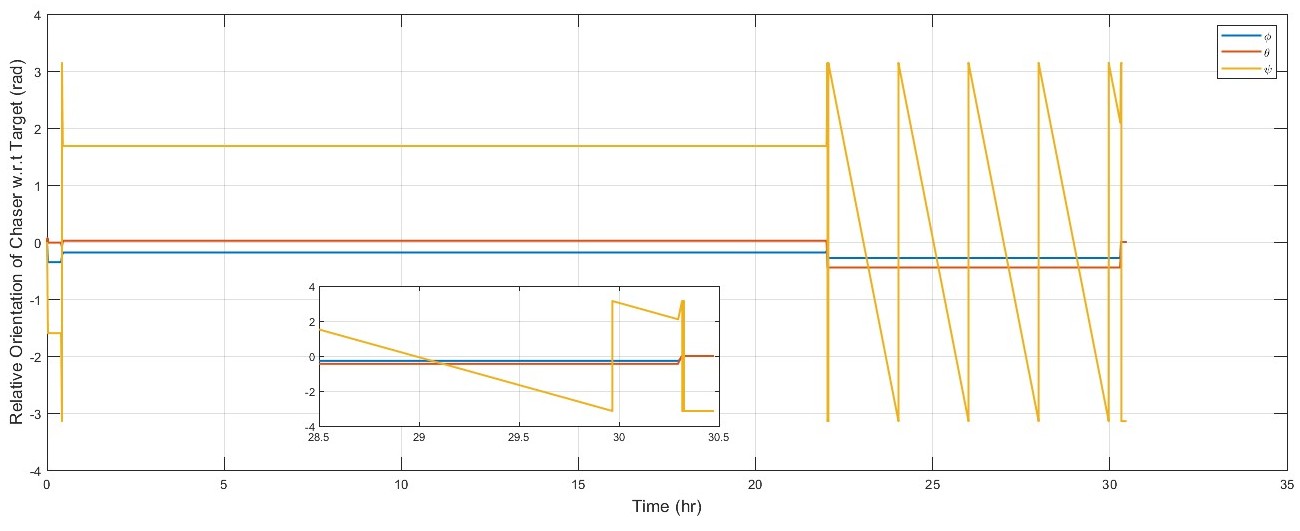}
    \caption{Relative orientation performed by chaser satellite to dock with the target satellite}
    \label{rel_orient}
\end{figure}
\section{Results and Discussion}
The mass of both the satellite are considered to be $1000\,kg$. And the inertia matrix of both satellites are taken as $\mathbf{J_t}\,=\,diag(500,2500,2500)\,\text{kg-m}^2$. The initial classical orbital elements of the chaser and target satellite is given in Table \ref{initial_states_table}. The three-dimensional trajectory of chaser and target satellite during rendezvous and docking maneuver is shown in Fig. \ref{3d_trajectory}. \\
The initial relative position of the chaser satellite w.r.t the target satellite is specified as  $10,000\,km$. This marks the starting point of the rendezvous and docking mission, where the primary objective is to bring the chaser satellite into close proximity with the target satellite in a controlled and precise manner. The rendezvous condition is achieved when the radial distance between the chaser and target satellite reduces to \(1\,km\) which occurs at a mission duration of $29 \, hr \, 24 \, min$ as illustrated in Fig. \ref{rel_pos}. From \(10,000\,km\) to \(1\,km\), we have performed impulse maneuver using sliding mode controller. At this point the chaser satellite transitions into the docking phase, which is executed in two distinct phases to ensure safety and precision while avoiding any potential impact at the conclusion of the maneuver.

In the initial phase of the docking operation, the chaser satellite maintains a constant relative translational velocity of \(0.3\ m/s\)  w.r.t to the target satellite until the relative position is reduced to \(10\ m\). This phase ensures a smooth and predictable approach trajectory. Upon reaching the threshold relative position of $10 m$, the operation transitions to the second phase, wherein the relative translational velocity is progressively decreased to $0.03 m/s$ to enable precise and safe docking. This reduction in velocity is critical for achieving a soft and precise docking, minimizing any risk of collision or damage to either satellite.  The relative velocity during entire docking process is depicted in Fig. \ref{rel_vel}, highlighting the controlled deceleration during both the phases. Successful docking is confirmed after a total mission duration of \(30\ hr\ 28\min\ 29\ sec\), with the final separation distance between the centers of mass (c.m.) of the chaser and target satellites being $1\,m$ denoting the offset of docking port from the center of mass of the satellite. 
\begin{figure}
    \centering
    \includegraphics[width=1\linewidth]{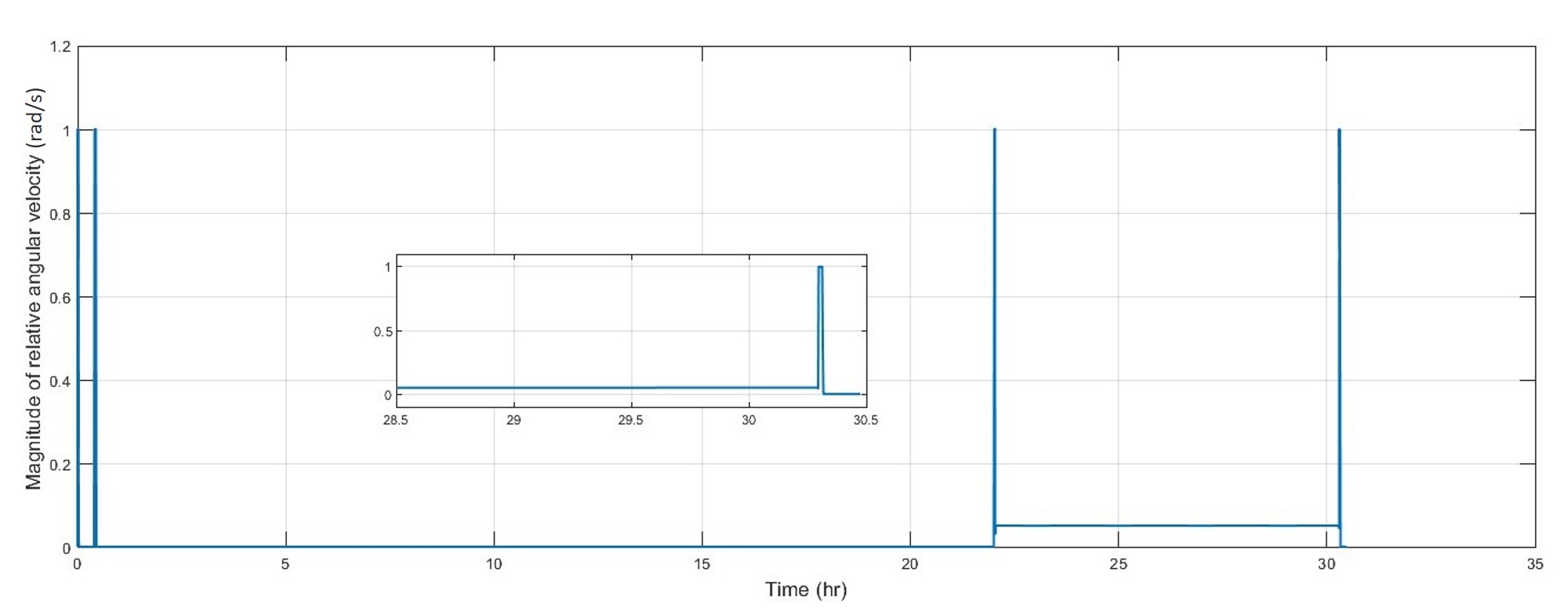}
    \caption{Relative angular velocity of chaser satellite with respect to the target satellite}
    \label{rel_ang_vel}
\end{figure}\\
\begin{table}[h]
    \centering
    \caption{Initial orbital elements of the chaser and target satellite }
    \begin{tabular}{@{} l l @{} l l @{}}
        \hline
        \textbf{Parameters} & \textbf{Chaser} & \textbf{Target}\\
        \hline
        Semi major axis (km)& $7500$& $8000$\\
        Eccentricity& $0.001$\qquad\qquad\qquad\quad& $0.0005$\qquad\\
        Inclination (\(^o\ deg\))& $30.1$& $30$\\
        RAAN (\(^o\ deg\))& $60.1$& $60$\\
        Argument of Periapsis (\(deg\))\qquad\quad\qquad& $120$& $120$\\
        True anomaly (\(^o\ deg\))& $30$& $310$\\
        \hline
    \end{tabular}
    \label{initial_states_table}
\end{table}
The optimization of mission results based on on constraints outlined in Section III results in a total impulse \(\Delta V\) of \(237.0428\ m/s\), which represents the cumulative velocity change required for the chaser satellite to complete its maneuver. The maximum burn duration for any single thruster firing is recorded to be \(4.9\ ms\), reflecting the high precision and efficiency of the propulsion system. In addition to translational motion, the mission also involves careful management of the chaser satellite's orientation.  The relative orientation of the chaser satellite with respect to the docking port of the target satellite was \([0,0,0]^T\) radians, and the relative angular velocity was \([0,0,0]^T\) rad/s as shown in Fig. \ref{rel_orient} and Fig. \ref{rel_ang_vel}, respectively. By the end of the docking phase, the chaser satellite must achieve a relative orientation of \((\pi,\pi,0)\) radians with respect to the target satellite. To accomplish this, the chaser satellite performs a rotational maneuver involving Euler angles \((\phi,\theta,\psi)=(0,0,-\pi)\)radians, as observed at the end of simulation in Fig. \ref{rel_orient}. 

\begin{figure}
    \centering
    \includegraphics[width=1\linewidth]{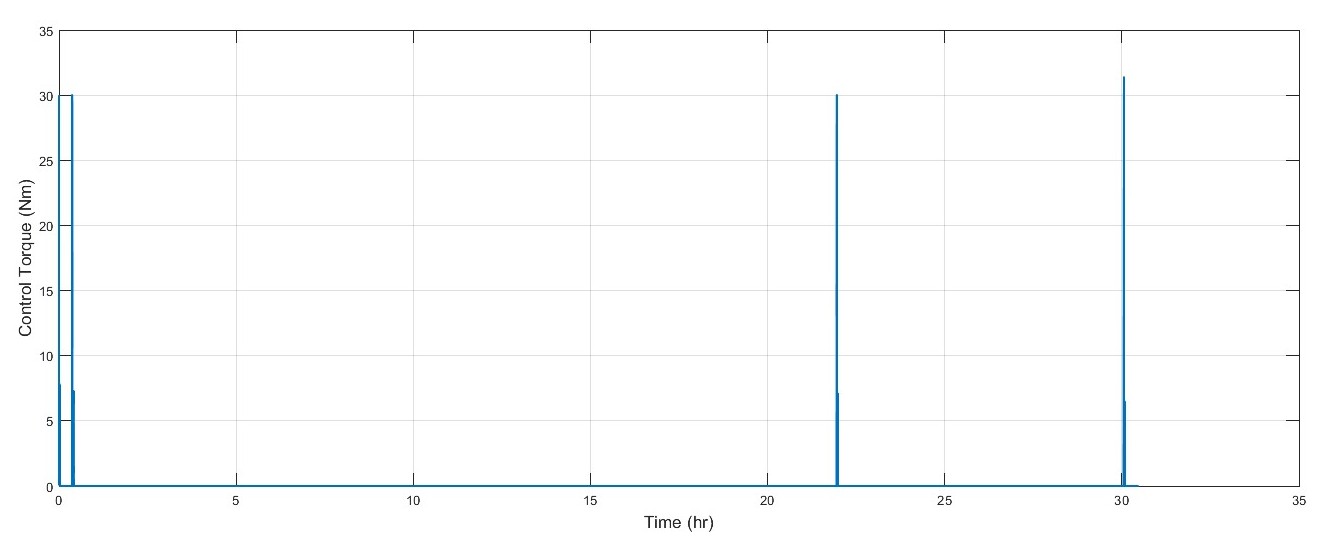}
    \caption{Control torque acting on the chaser satellite}
    \label{control_torque}
\end{figure}

Notably, the rendezvous condition is met within \(22\ hrs\), during which the relative velocity is successfully reduced to 1\textit{km}/\textit{s}, as demonstrated in Fig. \ref{rel_pos} and Fig. \ref{rel_vel}.  This highlights the effectiveness of the trajectory planning and control algorithms employed during the mission. Moreover, the control force and torque applied to the chaser satellite to achieve long range rendezvous and docking maneuver is shown in Fig. \ref{control_force} and Fig. \ref{control_torque}, respectively. The combination of precise velocity control, optimized thrust maneuvers, and accurate rotational adjustments ensures that the chaser satellite achieves its objectives safely and efficiently, culminating in a successful docking operation. 

\section{Conclusion}
The proposed work introduces a cross-feedback sliding mode controller for satellite rendezvous and docking (RvD) by interdependently regulating translational and rotational motions of the chaser satellite i.e., utilizing rotational state data to inform translational control laws and vice versa. The proposed work can be effectively applied in autonomous satellite servicing, on-orbit refueling, modular space station assembly and docking in planetary exploration missions. \\

\vspace{12pt}
\color{red}

\end{document}